%% file: main.tex
\documentclass[sigconf,balance,authorversion=true]{acmart}

\usepackage{xspace}
\usepackage[inline]{enumitem}
\usepackage{amsmath}
\usepackage{cleveref}
\usepackage{csquotes}
\usepackage{fontawesome}
\usepackage{tikz}
\usetikzlibrary{positioning, arrows, arrows.meta}

\copyrightyear{2021}
\acmYear{2021}
\setcopyright{acmlicensed}
\acmConference[SAC '21]{The 36th ACM/SIGAPP Symposium on Applied Computing}{March 22--26, 2021}{Virtual Event, Republic of Korea}
\acmBooktitle{The 36th ACM/SIGAPP Symposium on Applied Computing (SAC '21), March 22--26, 2021, Virtual Event, Republic of Korea}
\acmPrice{15.00}
\acmDOI{10.1145/3412841.3441997}
\acmISBN{978-1-4503-8104-8/21/03}

\newcommand{\Eg}{E.g.,\xspace}
\newcommand{\eg}{e.g.,\xspace}
\newcommand{\Ie}{I.e.,\xspace}
\newcommand{\ie}{i.e.,\xspace}
\newcommand{\peqes}{\textsc{PeQES}\xspace}

\definecolor{uulm}{RGB}{125,154,170}
\definecolor{uulm-akzent}{RGB}{169,162,141}
\definecolor{uulm-in}{RGB}{163,38,56}
\definecolor{uulm-med}{RGB}{38,84,124}
\definecolor{uulm-mawi}{RGB}{86,170,28}
\definecolor{uulm-nawi}{RGB}{189,96,5}

\usepackage{resources/jslistings}

\patchcmd{\thebibliography}{\clubpenalty4000}{\clubpenalty10000}{}{} % no orphans
\patchcmd{\thebibliography}{\widowpenalty4000}{\widowpenalty10000}{}{} % no widows
\patchcmd{\bibsetup}{\interlinepenalty=5000}{\interlinepenalty=10000}{}{} % no break of entry

\begin{document}

\title[PeQES: A Platform for Privacy-enhanced Quantitative Empirical Studies]{\texorpdfstring{PeQES: A Platform for Privacy-enhanced\\ Quantitative Empirical Studies}{PeQES: A Platform for Privacy-enhanced Quantitative Empirical Studies}}

\author{Dominik Mei{\ss}ner}
\orcid{0000-0002-2937-6306}
\email{dominik.meissner@uni-ulm.de}
\affiliation{%
  \department{Institute of Distributed Systems}
  \institution{Ulm University, Germany}
  \streetaddress{Albert-Einstein-Allee 11}
  \postcode{89081}
  \city{}
  \country{}
}

\author{Felix Engelmann}
\orcid{0000-0001-9356-0231}
\email{fe-research@nlogn.org}
\affiliation{%
  \department{Institute of Distributed Systems}
  \institution{Ulm University, Germany}
  \streetaddress{Albert-Einstein-Allee 11}
  \postcode{89081}
  \city{}
  \country{}
}

\author{Frank Kargl}
\orcid{0000-0003-3800-8369}
\email{frank.kargl@uni-ulm.de}
\affiliation{%
  \department{Institute of Distributed Systems}
  \institution{Ulm University, Germany}
  \streetaddress{Albert-Einstein-Allee 11}
  \postcode{89081}
  \city{}
  \country{}
}

\author{Benjamin Erb}
\orcid{0000-0002-5432-4989}
\email{benjamin.erb@uni-ulm.de}
\affiliation{%
  \department{Institute of Distributed Systems}
  \institution{Ulm University, Germany}
  \streetaddress{Albert-Einstein-Allee 11}
  \postcode{89081}
  \city{}
  \country{}
}

\begin{abstract}
  Empirical sciences and in particular psychology suffer a methodological crisis due to the non-reproducibility of results, and in rare cases, questionable research practices.
  Pre-registered studies and the publication of raw data sets have emerged as effective countermeasures.
  However, this approach represents only a conceptual procedure and may in some cases exacerbate privacy issues associated with data publications.
  We establish a novel, privacy-enhanced workflow for pre-registered studies.
  We also introduce \peqes, a corresponding platform that technically enforces the appropriate execution while at the same time protecting the participants' data from unauthorized use or data repurposing.
  Our \peqes prototype proves the overall feasibility of our privacy-enhanced workflow while introducing only a negligible performance overhead for data acquisition and data analysis of an actual study.
  Using trusted computing mechanisms, \peqes is the first platform to enable privacy-enhanced studies, to ensure the integrity of study protocols, and to safeguard the confidentiality of participants' data at the same time.
\end{abstract}




\maketitle

\input{sections/1-introduction.tex}
\input{sections/2-background.tex}
\input{sections/3-workflow.tex}
\input{sections/4-prototype.tex}
\input{sections/5-relatedwork.tex}
\input{sections/6-discussion.tex}
\input{sections/7-conclusion.tex}

\bibliographystyle{ACM-Reference-Format}
\bibliography{references}

\end{document}

%% file: sections/1-introduction.tex
%!TeX spellcheck = en-US
\section{Introduction}
\label{sec:intro}

Empirical research disciplines such as psychology rely on the use of surveys to assess phenomena that are not directly measurable.
Appropriate operationalizations enable statistical analyses of these phenomena, \eg through survey items.
For instance, a clinical psychological study might evaluate the outcome and efficiency of a therapeutical intervention by comparing self-reports of subjects before and after the treatment.
Similarly, a socio-psychological study might explore the association of personality characteristics with certain behavioral tendencies using the correlation of operationalized values.

However, empirical research disciplines\,---\,and in particular psychology\,---\,are increasingly facing two challenges in the realization and evaluation of studies and experiments, namely issues of the research process and privacy issues:
\begin{enumerate*}[label=(\arabic*)]
\item Non-reproducible study outcomes~\cite{Shrout2018} (\ie replication crisis) as well as questionable work methods such as HARKing~\cite{Kerr1998} (\ie post-hoc hypothesizing) or p-hacking~\cite{Head2015} (\ie data dredging for significant outcomes) are raising the question of the extent to which research results are conclusive;
\item Privacy requirements for empirical researchers are becoming more demanding, particularly due to the increasing use of new technology in their research~\cite{Kargl2019}.
\end{enumerate*}
\Eg studies that utilize mobile sensing differ almost completely from traditional pen-and-paper surveys in terms of data collection, data handling, and data processing.

Pre-registration has now been suggested to tackle the former challenge: the registration, publication, and review of a study protocol and hypotheses prior to the actual data collection and analysis~\cite{Shrout2018}.
This practice is sometimes complemented with the publication of primary data sets to verify analyses and to foster re-use~\cite{OSF2015}.
This practice yields new privacy issues~\cite{Boronow2020}, though (\eg de-anonymization and re-identification attacks against study participants).

To overcome both challenges at once for online studies, we suggest \peqes\,---\,a unified platform that embraces the conceptual idea of pre-registered studies, provides a technical runtime for empirical study procedures, ensures the integrity of the study protocols, and enforces the protection of the participants' privacy.
To introduce \peqes, we first discuss a general process of empirical studies and illustrate potential threats for data privacy and study integrity in \Cref{sec:background}.
In \Cref{sec:pew}, we introduce our privacy-enhanced workflow which takes these threats into account.
Next in \Cref{sec:prototype}, we demonstrate our platform prototype that implements the privacy-enhanced workflow.
\Cref{sec:relatedwork} compares our solution to existing approaches.
Finally, we discuss the implications of our \peqes platform in \Cref{sec:discussion} and summarize our contributions in \Cref{sec:conclusion}.

%% file: sections/2-background.tex
%!TeX spellcheck = en-US
\section{Quantitative Empirical Studies}
\label{sec:background}

Empirical research in psychology can be separated into quantitative approaches (\eg surveys with standardized self-report inventories) and qualitative approaches (\eg unstructured interviews).
Although both approaches are based on direct or indirect observations, only the quantitative method is primarily founded in mathematical modeling and necessitates statistical analysis.
For the remainder of this work, we only focus on quantitative psychological research that is conducted with surveys.
This covers both correlation studies and experimental study designs.
While our ideas can also be applied to fields other than psychology, we believe that strong privacy protection is most needed where highly sensitive personal data are processed.

\subsection{Research Methodology of Empirical Studies}

A streamlined process for quantitative psychological studies covers the following steps.
\begin{enumerate*}[label=(S\arabic*:)]
	\item Based on an overarching research question, related work, and available information and resources, the \emph{researchers} form a hypothesis;
	\item The \emph{researchers} design a study that addresses the hypothesis. This includes the choice of appropriate operationalizations, inventories, and variables for the phenomena, as well as the selection of the target sample.
	Experimental studies also require the choice of experimental manipulations or interventions;
	\item The \emph{ethics board} of the researchers' institution reviews the study design and approves or rejects it based on ethical aspects;
	\item The \emph{researchers} perform the actual study. This includes the recruitment of \emph{participants} and the collection of data;
	\item The analysis of the data allows the researchers to draw conclusions. In particular, statistical hypothesis testing determines whether the formulated hypothesis needs to be rejected or not, based on a certain significance level;
	\item By publishing the results, the \emph{researchers} can communicate their findings to their \emph{research community}.
\end{enumerate*}

Pre-registration extends the previous process by adding additional steps~\cite{OSF2015}:
\begin{enumerate*}[label=(P\arabic*:)]
	\item After S3, but before S4, the \emph{researchers} register the study by publishing their study design with peer reviewing;
	\item After S4, the \emph{researchers} document the actual course of the study and any deviations from the original design;
	\item Due to the prior acceptance of the study design, \emph{researchers} can now assume to publish their results in S6\,---\,even in case of negative outcomes\,---\,as long as they adhered to the original study protocol.
	 The publication step often requires the release of the raw data sets and corresponding analysis code.
\end{enumerate*}

\subsection{Involved Parties}
\label{sec:parties}

Next, we take a look at the different parties that are involved in this research process, consider their motivations, and hint to potential threats, if we assume malicious actors.

\begin{description}[leftmargin=8pt]
    \item[Researchers] %
    advance the state of knowledge in their discipline by designing and conducting studies and by publishing the outcomes of their studies.
    Therefore, researchers follow principles and best practices of their discipline, apply rigorous methods, and work honestly.

    \item[Ethics Boards] supervise the ethical integrity of any study conducted by researchers affiliated with their institutions.
    An ethics board is a trustworthy body of an institution as it guarantees participants their safety, integrity, and privacy in the setting of the study.
    The board wants to protect the reputation of the institution and prevent scientific misconduct.

    \item[Participants] take part in studies, either by their own initiative, or because of participation rewards (\eg crediting of subject hours, monetary rewards).
    
    \item[Dishonest Participants] (so-called lurkers) might want to receive the reward without actual effort.
    This can lead to careless, haphazard, or random survey responses which reduce the data quality. Repeated participation by the same person represents another potential threat in which the attacker exploits the reward mechanism (so-called sybil attacks).

    \item[Research Communities] believe in the growth in knowledge in their disciplines.
    Therefore, a community critically reviews and evaluates studies, analyses, and outcomes published by their researchers.
    A community promotes good scientific practices and condemns scientific misconduct of its members.

    \item[Malicious Researchers] are primarily interested in the successful publication of their work.
    Therefore, they deviate from honest researchers when it comes to scientific conduct.
    The playbook of malicious researchers includes the following techniques:

	\begin{enumerate*}[label=(\arabic*)]
		\item Application of ethically questionable methods, either by skipping the ethics boards\footnote{As most journals in psychology, medicine, and social sciences require approvals by local institutional review boards, this technique also requires a forged approval.} (skipping S3), or by deliberate deviation from the original study design in S4;
		\item HARKing: exploring the data set after data collection in order to generate promising hypotheses, then using the very same data sets for statistical hypothesis testing. In this technique, the malicious researchers either ignore the general research methodology, or they modify their hypotheses post-hoc during S4;
		\item p-hacking: \eg exploring only subsets of the collected data sets to boost significance of outcomes. Here, the malicious researchers deliberately remove \enquote{outliers} in S4 to construct their desired outcome;
		\item The manipulation of data sets or the generation of synthetic data rows during S4 represents an even more severe variant.
	\end{enumerate*}

	Another attack vector for malicious researchers emerges through the potential linkage of data rows and participants. This enables malicious researchers to de-anonymize participants and explore individual responses, \eg for personal purposes.

    \item[Data Sniffers] access the publicly available data sets of studies with malicious intents.
    They try to link individual data rows to individual subjects using de-anonymization and re-identification techniques.
    Data sniffers take advantage of contextual information about the execution of the study (\eg recruitment procedure, time, and location).
    Furthermore, they use additional, external data sources (\eg social media profiles) for matching and linking.
\end{description}

%% file: sections/3-workflow.tex
%!TeX spellcheck = en-US
%

\begin{figure*}[t]
  \centering
  \input{resources/workflow.tex}
  \caption{The privacy-enhanced workflow for quantitative empirical studies.}
  \label{fig:workflow}
\end{figure*}
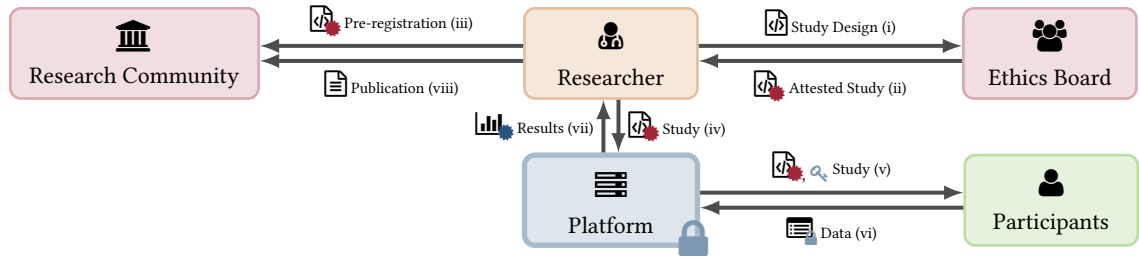

\section{Privacy-enhanced Study Workflow}
\label{sec:pew}

We propose an enhanced workflow for pre-registered studies in which the correctness of the procedure, the integrity of the study design, and the confidentiality of participant responses is protected in a technical manner.

First, our workflow requires a structured specification of the study design and the associated data analysis.
Second, the workflow introduces an institutional platform that handles the implementation of a study design and the actual execution of the study, namely the data collection and the statistical data analysis.
Technically, the platform represents a trusted execution environment.
Third, the workflow empowers certain parties to sign or verify study specifications or to encrypt data sets.

As shown in \Cref{fig:workflow}, the resulting workflow consists of eight consecutive steps:
\begin{enumerate}[label=(\roman*)]
    \item Based on a hypothesis (S1), the researchers specify a suitable study design (S2).
    This includes the specification of the study procedures and survey design as well as a pre-defined script for the statistical data analysis.
    The researchers then send the study design to their ethics board for approval.
    \item The ethics board reviews the submitted study by considering the structured specification (S3).
    Furthermore, the ethics board audits the intended analysis and assesses its privacy impact for participants (\eg risk of re-identifiable records in the analysis output).
    Upon approval, the ethics board cryptographically signs the study design.
    \item After approval, the researchers pre-register the study and present it to the research community (P1). This also includes the publication of the study specification so that other researchers can check the analysis in use.
    Additionally, the research community can use its specific domain knowledge to peer review the analysis code and complement the ethics board's assessment.
    \item Next, the researchers submit the study to the platform.
    The platform only accepts studies that have undergone prior acceptance by the ethics board.
    For valid studies, it creates corresponding survey forms based on the specification.
    \item Participants are intivited to take part in active studies (S4).
    Therefore, the platform provides them with information about the study and demonstrates them the approval of the ethics board.
    \item The survey responses of a participant are encrypted and securely transmitted to the platform for data collection.
    \item By a pre-defined trigger (\eg time period, total number of participations) or upon instruction of the researchers, the platform eventually transitions the study from the data collection phase to the data analysis phase.
    Therefore, the platform closes the data collection and then executes the pre-defined and signed analysis script exactly once.
    That is, the researchers only receive the final outputs of their pre-defined analyses (S5).
    However, they cannot explore the raw data sets, run interactive analyses, or make changes to their scripts.
    \item Drawing on the provided results, the researchers compile a publication of the study outcomes (S6/P3).
\end{enumerate}

%% file: resources/workflow.tex
\usetikzlibrary{calc, patterns, positioning}
\newcommand\icon[1]{{\large#1}}
\newcommand\badge[3]{\icon{#1}\hspace{-.5em}\raisebox{-.4em}{\color{#2}\footnotesize#3}}
\newcommand\blockicon[1]{{\LARGE#1}\\[5pt]}
\newcommand\block[5]{
    \node[block, fill=#1!15, draw=#1!40, #2] (#3) {};
    \node[yshift=.3cm] at (#3.south) {#5};
    \node[yshift=.85cm] at (#3.south) {{\LARGE#4}};
}

\begin{tikzpicture}[
    block/.style={
        fill=uulm-med!15,
        draw=uulm-med!40,
        line width=1pt,
        rounded corners,
        inner sep=.25cm,
        text width=3.3cm,
        minimum height=1.2cm,
        inner sep=0,
        align=center
    },
    arrow/.style={
        -latex,
        line width=1.5pt,
        draw=black!70
    }
]

    \block{uulm-med}{text width=2.3cm, line width=2pt}{platform}{\faServer}{Platform}
    \node[above left=-.7em and -.7em of platform.south east] {\color{uulm-med!60}\Huge\faLock};

    \block{uulm-nawi}{above=.7cm of platform, text width=2.3cm}{researcher}{\faUserMd}{Researcher}
    \block{uulm-in}{right=3.5cm of researcher, text width=2.3cm}{ethicsboard}{\faUsers}{Ethics Board}
    \block{uulm-in}{left=3.5cm of researcher}{community}{\faUniversity}{Research Community}
    \block{uulm-mawi}{right=3.5cm of platform, text width=2.3cm}{participants}{\faUser}{Participants}

    \draw[arrow] ($ (researcher.east) + (0, .1) $) -- node[midway,above] {\scriptsize\icon{\faFileCodeO} Study Design (i)} ($ (ethicsboard.west) + (0, .1) $);
    \draw[arrow] ($ (ethicsboard.west) + (0, -.1) $) -- node[midway,below] {\scriptsize\badge{\faFileCodeO}{uulm-in}{\faCertificate} Attested Study (ii)} ($ (researcher.east) + (0, -.1) $);
    \draw[arrow] ($ (researcher.west) + (0, .1) $) -- node[midway,above] {\scriptsize\badge{\faFileCodeO}{uulm-in}{\faCertificate} Pre-registration (iii)} ($ (community.east) + (0, .1) $);

    \draw[arrow] ($ (researcher.south) + (.1, 0) $) -- node[midway,right] {\scriptsize\badge{\faFileCodeO}{uulm-in}{\faCertificate} Study (iv)} ($ (platform.north) + (.1, 0) $);
    \draw[arrow] ($ (platform.east) + (0, .1) $) -- node[midway,above] {\scriptsize\badge{\faFileCodeO}{uulm-in}{\faCertificate}\badge{~}{uulm-med!60}{\textcolor{black}{,} \faKey} Study (v)} ($ (participants.west) + (0, .1) $);
    \draw[arrow] ($ (participants.west) + (0, -.1) $) -- node[midway,below] {\scriptsize\badge{\faListAlt}{uulm-med!60}{\faLock} Data (vi)} ($ (platform.east) + (0, -.1) $);

    \draw[arrow] ($ (platform.north) + (-.1, 0) $) -- node[midway,left] {\scriptsize\badge{\faBarChart}{uulm-med}{\faCertificate} Results (vii)} ($ (researcher.south) + (-.1, 0) $);
    \draw[arrow] ($ (researcher.west) + (0, -.1) $) -- node[midway,below] {\scriptsize\icon{\faFileTextO} Publication (viii)} ($ (community.east) + (0, -.1) $);
\end{tikzpicture}

%% file: sections/4-prototype.tex
%!TeX spellcheck = en-US
\section{The \peqes Platform}
\label{sec:prototype}

\newcommand{\Study}{\mathcal{S}}
\newcommand{\Researcher}{\mathcal{R}}
\newcommand{\EthicsBoard}{\mathcal{B}}
\newcommand{\Enclave}{\mathcal{E}}
\newcommand{\Participant}{\mathcal{P}}

\newcommand{\Enc}{\textsf{Enc}}
\newcommand{\Dec}{\textsf{Dec}}
\newcommand{\Ver}{\textsf{Ver}}
\newcommand{\Sig}{\textsf{Sig}}
\newcommand{\pk}{\textsf{pk}}
\newcommand{\sk}{\textsf{sk}}
\newcommand{\id}{\textsf{id}}
\newcommand{\SMK}{\textsf{SMK}}
\newcommand{\EK}{\textsf{EK}}
\newcommand{\DiHe}{\textsf{DH}}

\newcommand{\drawrand}{\stackrel{\$}{\leftarrow }}

In this section, we introduce our platform for privacy-enhanced quantitative empirical studies, or \peqes for short.
In our prototype, the platform is responsible for study management, participant data aggregation, and execution of the analysis script.
The platform has to be trustworthy for participants and researchers, as it receives the raw data of all participants and is tasked with computing the results of the statistical data.
There are multiple technical approaches to implement this kind of trusted service for privacy preserving data processing, such as secure multiparty computation, fully homomorphic encryption, and trusted execution environments.
Each of these approaches have different properties, advantages, and disadvantages, which we briefly discuss in \Cref{sec:relatedwork}.
For \peqes, we opted for a trusted execution environment.

\emph{Trusted execution environments} (TEE) are a promising technology to ensure data integrity and confidentiality, even if the underlying operating system is compromised.
A TEE may be implemented by \emph{hardware enclaves} that are based on a secure section of the processor with access to a protected memory area that is inaccessible to the OS.  This allows code running inside the enclave to keep secrets from the OS and other parts of the system.
I/O operations and networking is provided by the untrusted OS and has to be secured by other mechanisms.
Such enclaves usually allow remote parties to remotely attest the authenticity of the hardware as well as the code running inside the enclave, thus enabling secure and trusted communication with software inside the enclave.
To persistently store data, enclaves rely on the untrusted OS.
To ensure confidentiality and integrity of stored data, enclaves have the capability to derive sealing keys that are based on the currently executed software and the unique hardware.
Before passing data for storage to the OS, its authenticity and confidentiality can be assured through an authenticated encryption with a derived sealing key.
The stored data can only be decrypted by the same version of the platform running on the same processor.
We implemented a prototype of our platform using the widely available Intel Software Guard Extensions (Intel SGX)\footnote{\url{https://software.intel.com/en-us/sgx}}, an extension to the instruction set of Intel CPUs that enables the usage of secure enclaves.
As our \peqes protocols are weakly coupled to the enclave, it can easily be ported to a different TEE in the case of serious vulnerabilities in SGX.

\subsection{Architecture}

\begin{figure}[t]
    \centering

    \begin{tikzpicture}[arrow/.style={{Latex[length=1.1mm,width=2.1mm]}-{Latex[length=1.1mm,width=2.1mm]}, thick}]
        \node[fill=uulm-akzent!20,draw=uulm-akzent,thick,minimum width=5.6cm,text depth=2.4cm] at (0,0) (platform) {\peqes Platform};
        \node[anchor=north west, yshift=-.5cm, xshift=.1cm, fill=white, draw, thick, align=center, text width=1.7cm, text depth=.55cm] at (platform.north west) (datastore) {{\Huge\faDatabase}\\[5pt]Data Store};
        \node[below=.1cm of datastore, fill=white, draw, thick, align=center, text width=1.7cm, text depth=.4cm] (tls) {TLS\\ Termination};

        \node[anchor=north east, yshift=-.5cm, xshift=-.1cm, fill=uulm-med!20, draw=uulm-med, thick, minimum width=3.25cm, text depth=1.8cm] at (platform.north east) (enclave) {Trusted SGX Enclave};
        \node[anchor=north west, yshift=-.5cm, xshift=.1cm, fill=white, draw, thick, text width=1.15cm, minimum height=1.65cm, align=center] at (enclave.north west) (http) {HTTP Request Handler};
        \node[fill=white,draw,thick, text width=1.15cm,right=.3cm of http.north east, anchor=north west, align=center] (studies) {Studies};
        \node[fill=white,draw,thick, text width=1.15cm,below=.3cm of studies, align=center] (engine) {Statistics Engine};

        \node[below left=0cm and .4cm of platform.north west, fill=white, draw, thick, align=center, text width=2.2cm, minimum height=.5cm] (ias) {};
        \node[above left=0cm and .4cm of platform.south west, fill=uulm-nawi!20, draw, thick, align=center, text width=2.2cm, minimum height=.5cm] (researcher) {};
        \node[above=.1cm of researcher, fill=uulm-mawi!20, draw, thick, align=center, text width=2.2cm, minimum height=.5cm] (participant) {};
        \node[above=.1cm of participant, fill=uulm-in!20, draw, thick, align=center, text width=2.2cm, minimum height=.5cm] (ethicsboard) {};

        \node[xshift=.3cm] at (ias.west) {\faServer};
        \node[xshift=.3cm] at (researcher.west) {\faLaptop};
        \node[xshift=.3cm] at (participant.west) {\Large\faMobile};
        \node[xshift=.3cm] at (ethicsboard.west) {\faLaptop};
        \node[anchor=west,xshift=.5cm] at (ias.west) {Intel IAS};
        \node[anchor=west,xshift=.5cm] at (researcher.west) {Researcher};
        \node[anchor=west,xshift=.5cm,yshift=-.04cm] at (participant.west) {Participant};
        \node[anchor=west,xshift=.5cm] at (ethicsboard.west) {Ethics Board};

        \draw[arrow] (studies) -- (engine);
        \draw[arrow] (http.east|-studies.west) -- (studies.west);

        \draw[arrow] ([yshift=-2mm]datastore.east) -- ([yshift=-2mm]http.west|-datastore.east);
        \draw[arrow] (tls.east) -- (http.west|-tls.east);

        \draw[arrow] (researcher.east) -- (tls.west|-researcher.east);
        \draw[arrow] (participant.east) -| ([xshift=-3mm]tls.west) -- (tls.west);
        \draw[arrow] (ethicsboard.east) -| ([yshift=-1.5mm,xshift=-2mm]tls.north west) -- ([yshift=-1.5mm]tls.north west);
        \draw[arrow] (ethicsboard.north) -- (ias);
    \end{tikzpicture}

    \caption{\peqes prototype architecture.\label{fig:architecture}}
\end{figure}
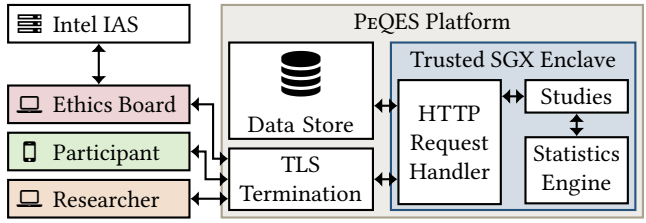

Our prototypical realization of the \peqes platform is depicted in \Cref{fig:architecture}.
The platform provides an HTTP-based API that allows different parties (researcher, ethics board, participants) to interface with it.
To keep the participation hurdle as low as possible, we assume that participants do not wish to install dedicated software to participate and instead use a modern web browser.
The HTTP communication applies common TLS transport encryption, which is terminated outside the hardware enclave.

We argue that moving the TLS termination inside the enclave would not be sufficient to establish trust in the enclave, without performing a remote attestation during the TLS handshake or pinning a public key for the enclave.
The platform can simply eclipse the enclave and obtain another valid certificate for the domain.
As we assume the clients to be standard web browsers, a remote attestation during the TLS handshake is not feasible.
Neither the deprecated HTTP Public Key Pinning~\cite{RFC7469}, nor its replacement Certificate Transparency~\cite{RFC6962} enable a client to differentiate a malicious certificate issued by the platform from a legitimate certificate used by the enclave, as both are valid.

Instead, \peqes is built on transitive trust\,---\,participants trust the ethics board to work diligently, attesting the enclave, and therefore trust approved studies.

\subsection{Threat Model}

As generally discussed in \Cref{sec:parties}, we map the entities to the following security model:
While most \textbf{researchers $(\Researcher)$} have noble intentions, nevertheless we model them as fully malicious, trying to deviate from the protocol, \eg to gain survey information or alter their analysis. We assume that they all perform pre-registered studies and the study design itself is considered public.
The \textbf{ethics board $(\EthicsBoard)$} acts honestly in any aspect.
As we require honest answers to the study, the \textbf{participants} $(\Participant)$ have to act honestly.
Sybil attacks may be dealt with by an authentication provider, which is out of scope.
Dubious responses by lurkers (\eg always selecting the first option) are expected to be excluded as part of the actual data analysis phase (\ie data cleansing).
The \textbf{platform provider} is modeled as covert, i.e. tries to extract information and only deviates from the protocol, if it is unlikely to be caught.
It is trusted with the distribution of the correct web application to the participants but is interested in extracting raw survey results.
The platform does not engage in a denial-of-service attack against the enclave.
The \textbf{secure enclave $(\Enclave)$} is assumed to be secure and unbreachable, and the code inside attested.
Attacks on the confidentiality of the SGX enclave and side-channel attacks are out of scope.
Data sniffers are modeled as any external malicious attacker, who aims to get information about the study answers.

\subsection{Protocol Flow}

\paragraph{Setup} To prepare our \peqes platform, the ethics board $\EthicsBoard$ generates a key pair $(\EthicsBoard_\sk,\EthicsBoard_\pk)$.
The public key $\EthicsBoard_\pk$ is published for verification of study approvals.
Multiple ethics boards may form a PKI-like structure.
The researcher $\Researcher$ generates a key-pair $(\Researcher_\sk,\Researcher_\pk)$.
In addition, a secure signature scheme $(\Sig,\Ver)$ and a symmetric encryption scheme $(\Enc,\Dec)$ is set up.

\paragraph{Register Study}
A researcher registers a new study $\Study$ by uploading its specification (\ie name, description, survey, analysis script) and the researcher's public key $\Researcher_{\pk}$ to an enclave $\Enclave$ of the \peqes platform. A signature $\Sig_{\Researcher_\sk}(\Study)$ attests the intent of $\Researcher$.
Once uploaded, $\Enclave$ verifies the signature and generates a study key-pair $(\Study_\sk, \Study_\pk)$ for the study, which is kept secure in the protected memory of the enclave.

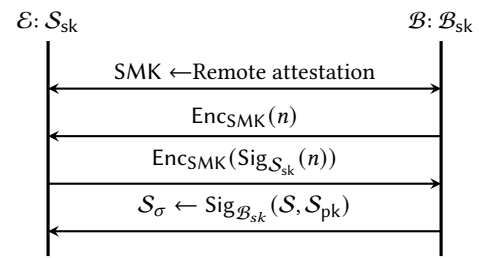
\begin{figure}[ht]
    \centering

    \begin{tikzpicture}[>=stealth]
        \node[minimum height=9em, label=above:{$\Enclave$: $\Study_\sk$}] (a) {};
        \node[minimum height=9em, right=5cm of a, label=above:{$\EthicsBoard$: $\EthicsBoard_\sk$}] (b) {};
        \draw[very thick] (a.north) -- (a.south);
        \draw[very thick] (b.north) -- (b.south);

        \draw[<->,thick] ([yshift=-2em] a.north) -- node[above]{$\SMK \leftarrow $Remote attestation} ([yshift=-2em] b.north);
        \draw[<-,thick] ([yshift=-4em] a.north) -- node[above]{$\Enc_{\SMK}(n)$} ([yshift=-4em] b.north);
        \draw[->,thick] ([yshift=-6em] a.north) -- node[above]{$\Enc_{\SMK}(\Sig_{\Study_\sk}(n))$} ([yshift=-6em] b.north);
        \draw[<-,thick] ([yshift=-8em] a.north) -- node[above]{$\Study_\sigma \leftarrow \Sig_{\EthicsBoard_{sk}}(\Study, \Study_\pk)$} ([yshift=-8em] b.north);
    \end{tikzpicture}

    \caption{Protocol flow for approving a study.}
    \label{fig:approve}
\end{figure}

\paragraph{Approve Study}
To approve a study $\Study$ (\Cref{fig:approve}), the ethics board $\EthicsBoard$ performs a remote attestation to establish a secure channel with the platform enclave $\Enclave$ through a shared master key $(\SMK)$.
During the attestation, $\EthicsBoard$ verifies that $\Enclave$ is executed in a genuine trusted execution environment and that the platform software has not been tampered with.
$\EthicsBoard$ uses the secure channel to verify that $\Study_\sk$ is only known to $\Enclave$ through a combination of the assurances of the remote attestation and a proof of knowledge (\ie signing a nonce $n$, provided by $\EthicsBoard$, with $\Study_\sk$).
If successful and if $\Study$ is compliant, $\EthicsBoard$ signs $(\Study_\sigma = \Sig_{\EthicsBoard_\sk}(\Study,\Study_\pk))$ the study as approved.

\begin{figure}[ht]
    \centering

    \begin{tikzpicture}[>=stealth]
        \node[minimum height=11em, label=above:{$\Enclave$: $\Study_\sk$}] (a) {};
        \node[minimum height=11em, right=5cm of a, label=above:{$\Participant$: $\EthicsBoard_\pk$}] (b) {};
        \draw[very thick] (a.north) -- (a.south);
        \draw[very thick] (b.north) -- (b.south);

        \draw[->,thick] ([yshift=-2em] a.north) -- node[above]{$\Study,\Study_\pk,\Study_\sigma$} ([yshift=-2em] b.north);
        \node[below=2.5em and .1em of b.north, fill=white] {$\Ver_{\EthicsBoard_\pk}(\Study_\sigma,\Study)$};
        \node[below=2.5em and .1em of a.north, fill=white] {gen $G_\Enclave$};
        \draw[->,thick] ([yshift=-5em] a.north) -- node[above]{$G_{\Enclave,\pk}, G_\sigma$} ([yshift=-5em] b.north);
        \node[below=5.5em and .1em of b.north, fill=white,xshift=-3pt,text width=2.5cm,align=center] {$\Ver_{\Study_\pk}(G_{\sigma},G_{\Enclave,\pk})$,\\gen $G_\Participant$, $\DiHe(G_\Participant,G_\Enclave)$};
        \draw[<-,thick] ([yshift=-10em] a.north) -- node[above, near start]{$G_{\Participant,\pk}, \Enc_{\EK}(\Study_\Participant)$} ([yshift=-10em] b.north);
    \end{tikzpicture}

    \caption{Protocol flow for participating in a study.}
    \label{fig:participate}
\end{figure}
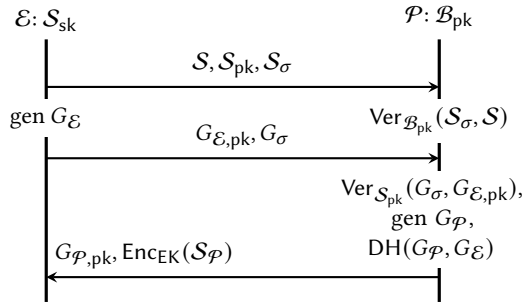

\paragraph{Study Participation}
To participate in the study (\Cref{fig:participate}), the participant $\Participant$ fetches $\Study$, $\Study_\pk$, and $\Study_\sigma$ from the platform, together with a trusted JavaScript web application. They validate the approval of $\EthicsBoard$ by $\Ver_{\EthicsBoard_\pk}(\Study_\sigma,\Study)\stackrel{?}{=}1$ and abort on failure.

After completion of the study, $\Participant$ prepares to upload their response $\Study_\Participant$ for $\Study$.
First, $\Participant$ generates a temporary key pair $(G_{\Participant,\sk},$ $G_{\Participant,\pk})$ and establishes an ephemeral key $\EK$ with $\Enclave$ and its temporary key $(G_{\Enclave,\sk},$ $G_{\Enclave,\pk})$ over an authenticated Diffie-Hellman exchange $\DiHe(G_{\Enclave},G_\Participant)$.
\Ie the temporary public key $G_{\Enclave,\pk}$ of $\Enclave$ is signed $G_\sigma =\Sig_{\Study_\sk}(G_{\Enclave,\pk})$.
Then $\Participant$ verifies $\Ver_{\Study_\pk}(G_{\sigma},G_{\Enclave,\pk})$. By verifying $\Study_\sigma$ and $G_{\sigma}$, $\Participant$ can skip the remote attestation of $\Enclave$ during this step and still validate the enclave's authenticity by transitive trust and the assumption, that $\EthicsBoard$ performed a remote attestation.
After successful verification, the participant's response $\Study_\Participant$ is encrypted $\Enc_\EK(\Study_\Participant)$ and uploaded to $\Enclave$ and then processed there.

Next, $\Enclave$ uses $\EK$ to decrypt $\Dec_\EK(\Study_\Participant)$. Due to storage limitations, the results are persisted outside of $\Enclave$. They are encrypted with a symmetric key derived from $\Study_\sk$. Integrity and protection against roll-back attacks by the platform provider is achieved by storing the encrypted objects in a Merkle tree. Thereby the enclave merely stores the root hash of the tree and performs efficient updates on the data structure.

\paragraph{Study Completion}
To complete a study $\Study$, the researcher authenticates to $\Enclave$ with the help of $\Researcher_{sk}$.
If valid, $\Enclave$ stops accepting new participants and executes the predefined analysis script of $\Study$.
The result $\Study_\text{result}$ of the script is transferred to $\Researcher$ for publication.
Optionally, $\Enclave$ signs $\Sig_{\Study_\sk}(\Study_\text{result})$ for public acceptance of the results.
Optionally, if $\EthicsBoard$ mandates, the raw results of the study are deleted by $\Enclave$ so that it is impossible to run the evaluation script twice.

\subsection{Implementation}

We have implemented \peqes using the Rust-based Fortanix Enclave Development Platform\footnote{\url{https://edp.fortanix.com/}}.
The prototype implementation of the \peqes platform is available under an open source license\footnote{\url{https://github.com/vs-uulm/peqes/tree/sac21}}.
In total we implemented our platform prototype in 1459 lines of code.

\begin{figure}[ht]
    \centering

    \includegraphics[width=\columnwidth]{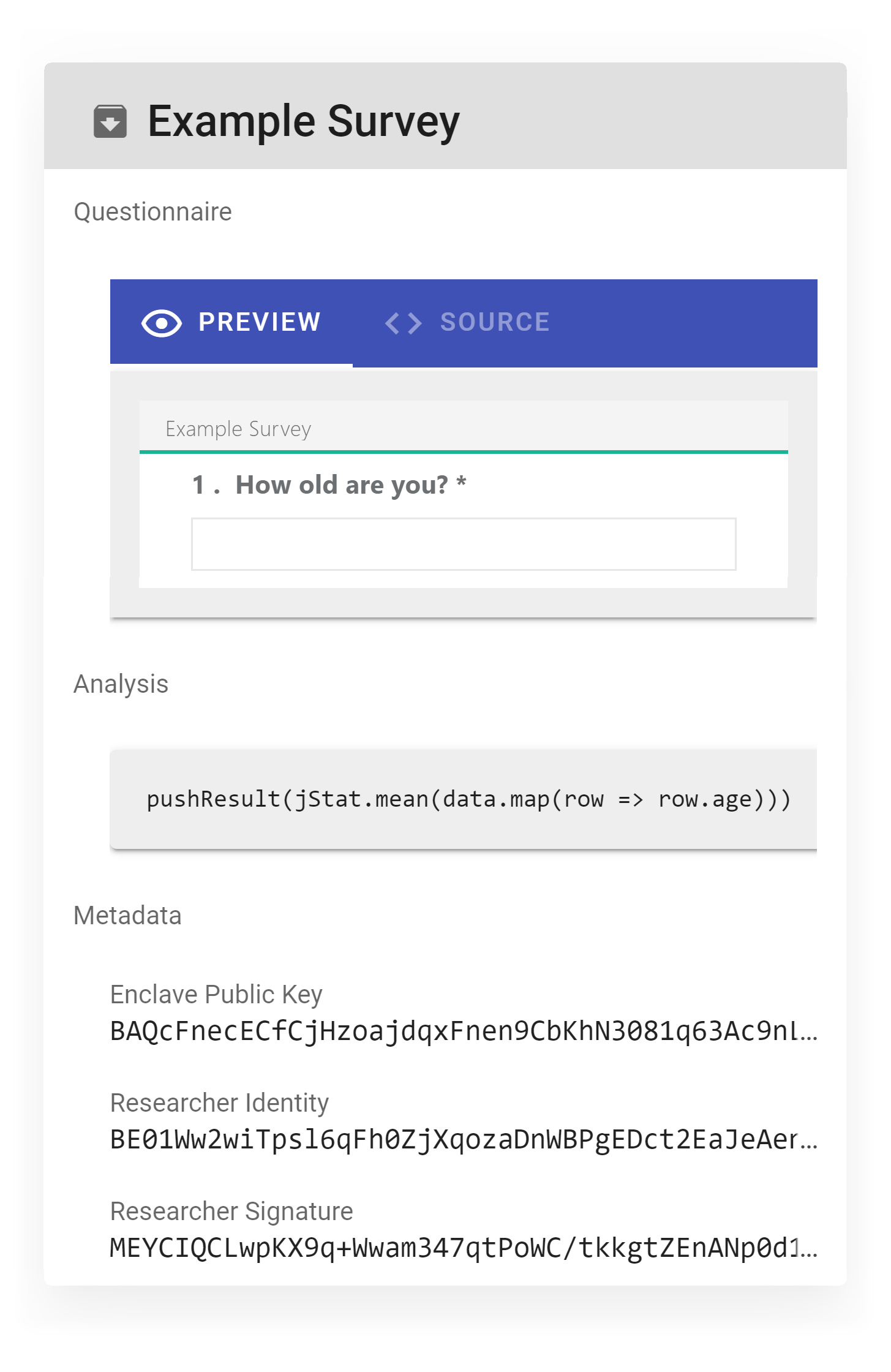}

    \caption{Study preview for the researcher and ethics board that with additional meta information to identify the researcher and enclave.}
    \label{fig:screenshot}
\end{figure}

Next, we implemented the client as a Vue.js web application in 871 lines of code (c.f., \Cref{fig:screenshot}).
It relies on native cryptography primitives provided by the Web\-Cryptography API~\cite{Watson2017} to encrypt participants' responses.
Survey specifications are represented as JSON and rendered using the SurveyJS library\footnote{\url{https://surveyjs.io/Overview/Library/}}.
The analysis scripts can be composed in a JavaScript-based DSL with access to the statistics library \texttt{jstat}.

The statistics component of the platform is based on the embedded JavaScript engine QuickJS\footnote{\url{https://bellard.org/quickjs/}} with custom Rust bindings.
The JavaScript interpreter is limited to standard JavaScript functions with the exception of statistics primitives from the \texttt{jstat} library\footnote{\url{https://jstat.github.io/}} that is embedded in the enclave.
Especially I/O access, such as accessing the file system or accessing network resources, is not possible to prevent obvious side channel attacks.
The memory of the engine is located in the encrypted section of the enclave.
The JavaScript environment is sandboxed and has only access to its internal state but not to other memory inside or outside the enclave.
Prior to execution, the input data is copied into the memory of engine in form of a JSON array.
The results computed by the analysis script are collected as a JSON object that is copied back to the enclave following the execution.

\subsection{Security Evaluation}
The \peqes platform is secure against our specified threat model, as the following measures are implemented.
The enclave $(\Enclave)$, participants $(\Participant)$, and ethics board $(\EthicsBoard)$ act honestly and therefore need no further consideration.
As the raw data of a study are not published, which they classically are for reproduction purposes, data sniffers have no access at all.

The platform provider is assumed to act covertly and tries to learn as much as possible from the interactions.
The delivery of the website including the cryptography code contains no sensitive information but allows to ship code to exfiltrate participant's data.
However this inflicts a prohibitive risk of getting caught and losing reputation.
During the \emph{Register Study} workflow, the platform provider is able to impersonate a legit enclave by emulation and get access to an already public study information.
In addition, such an emulation is detected by the ethics board during the remote attestation of the \emph{Approve Study} workflow.
Any further communication through the platform to the enclave is encrypted and authenticated by the signature of the ethics board.
This restricts the platform to merely collect meta-data of all connections without knowing the association to a specific study.
With storage inside enclaves limited, the platform provider acts as a storage device for the enclave.
As the data is authenticated and encrypted by the enclave, the only attack vector is to correlate encrypted communication with the size of the encrypted data.
This leaks at most, if a submitted answer is accepted and persisted by the enclave, which all of the submissions should be.

The malicious researchers are confined to legitimate actions by our protocol.
During \emph{Register Study}, a malicious research may upload a flawed study under the identity of someone else to harm reputation or cause a penalty.
This is prevented by the mandatory signature of the researcher requiring the secret key of the impersonated one.
The malicious act more difficult to detect is an untruthful study design.
A common goal is the extraction of participants' answers and the possibility to change the analysis retrospectively to align the study results with the hypothesis or improve the confidence.
Our platform includes the following countermeasures.
The participants' answers are stored encrypted by the enclave without access by anyone outside the enclave.
The analysis script provided can only be executed once, which is also enforced by the enclave.
This prevents data reconstruction from multiple samplings of random subsets or similar.
It remains to verify that the study itself complies with ethical regulations and common sense.
This includes that the results properly anonymize answers.
In \peqes, the ethics board is tasked to ensure these properties for all studies they approve.
The questionnaire part of the study is a finite set of questions and possible answers, similar to classical studies, and therefore easy to audit.

The analysis script remains as the most difficult part to verify if it adheres to a benevolent cause.
Not only is it a new component, many ethics boards are not prepared for, but even from a computer science perspective it is difficult to assess if the analysis will do only what is evident from looking at it.
Especially dynamic, interpreted environments allow for subtle side effects, difficult to find even after thorough inspection.
Even though the interpreter is sandboxed as described in the previous section, we acknowledge that a Turing-complete analysis/query language has inherent caveats.
If analyzing the script is not feasible for ethics boards, \peqes is modular by design and the statistics component can be adapted to provide an interpreter for a specifically restricted domain specific language (DSL) instead of the current JavaScript version.
Therefore, enabling a trade-off between functionality and computational expressiveness on the one hand, and security and reviewability on the other hand.
An example of such an alternative statistical component is a very limited flow-based DSL that only allows statistical primitives as pre-audited building blocks.
This enables easier composition of analysis scripts without coding experience and a more streamlined auditing process at the expense of limiting the feature set that is available for analysis scripts.

\subsection{Performance Evaluation}
We have conducted a set of experiments to compare the performance of the secure enclave of \peqes to a variant running without an enclave.
We have released all evaluation artifacts\footnote{\url{https://github.com/vs-uulm/peqes/tree/sac21}} under an open source license, following the Popper convention~\cite{Jimenez2017} for reproducible evaluations.
The experiments were carried out on two machines equipped with an Intel SGX capable Intel Core i7-7700 (quad-core with SMT; 3.60 GHz) CPU and 32 GB RAM, running Ubuntu 18.04.3 LTS, and connected using gigabit ethernet.

The sample study consisted of a short demographic questionnaire and BFI-10, a 10-item personality scale~\cite{Rammstedt2007}.
The analysis script explored differences in personality traits of two demographic sub-groups (split by age) using an independent two-sample \textit{t}-test (see Listing~\ref{lst:analysis}).
Using a synthetic workload generator on the second host, we sequentially submitted up to 10,000 participations to the platform for each evaluation run before transitioning to the the statistical analysis execution.
For each submission of a survey response, we measured the total response time at the client side.
For each study run, we also measured the total execution time of the analysis after collecting all responses.

\begin{figure}[ht]
    \centering
    \includegraphics{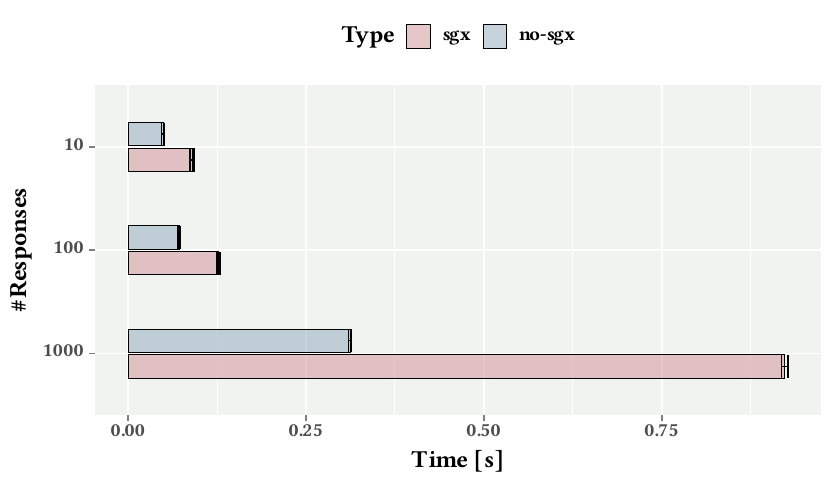}
    \caption{95\% confidence interval of the mean execution time of the statistical analysis as a function of platform implementation type and sample size based on 35 executions per variant.}
    \label{fig:comp-time}
\end{figure}

The unencrypted baseline implementation without enclave took an average of $M$=15.8 ms to handle a response submission ($P_{50}$=15.8 ms, $P_{99}$=19.8 ms).
In turn, \peqes required $M$=29.7 ms ($P_{50}$=29.6 ms, $P_{99}$=35.2 ms).
\Cref{fig:comp-time} illustrates the impact of different sample sizes ($N$=10 \dots 1,000) and the platform implementations (usage of SGX).
For a sample size of $N$=10,000 the mean computation time using SGX is 15.7 times as high as without SGX ($t_{\overline{\textrm{sgx}}}$=4.6 s, $t_{\textrm{sgx}}$=72.4 s).

\begin{lstfloat}[ht]
\begin{lstlisting}[language=JavaScript]
const s = [
  // demographic group with age <= 50
  { E: [], N: [], O: [], G: [], V: [] },
  // demographic group with age > 50
  { E: [], N: [], O: [], G: [], V: [] }
];

// iterate over participant responses
data.forEach((row) => {
  // recode items
  row['BFI'].item01 = 6 - row['BFI'].item01;
  row['BFI'].item03 = 6 - row['BFI'].item03;
  row['BFI'].item04 = 6 - row['BFI'].item04;
  row['BFI'].item05 = 6 - row['BFI'].item05;
  row['BFI'].item07 = 6 - row['BFI'].item07;

  // add response to respective sub demographic
  const g = row['age'] <= 50 ? 0 : 1;
  s[g].E.push(row['BFI'].item01, row['BFI'].item06);
  s[g].N.push(row['BFI'].item04, row['BFI'].item09);
  s[g].O.push(row['BFI'].item05, row['BFI'].item10);
  s[g].G.push(row['BFI'].item03, row['BFI'].item08);
  s[g].V.push(row['BFI'].item02, row['BFI'].item07);
});

// expose result object to platform
pushResult({
  E: ttest(s[0].E, s[1].E),
  N: ttest(s[0].N, s[1].N),
  O: ttest(s[0].O, s[1].O),
  G: ttest(s[0].G, s[1].G),
  V: ttest(s[0].V, s[1].V)
});
\end{lstlisting}
    \caption{Exemplified analysis script to compute an independent two-sample \textit{t}-test on the personality traits of two demographic sub-groups.\label{lst:analysis}}
\end{lstfloat}

All evaluation results indicate a noticeable performance degradation when using SGX.
However, we argue that the absolute values are completely within reasonable limits.
The increase of submission latency is hardly noticeable for participants and is far below the recommended limit of acceptable server response times of 200 ms\footnote{\url{https://developers.google.com/speed/docs/insights/Server}} by Google's project PageSpeed Insights.
The longer computation times become only apparent in larger samples sizes which are primarily used in specific sociological studies.
But even then the superlinear overhead is negligible as the computation is executed only once and lacks strict timing requirements.

\subsection{Limitations}
While our \peqes platform enables sound and privacy friendly quantitative studies, we acknowledge the following technical limitations. All communication is passed through the platform provider, which is responsible for TLS termination, hosting of the web applications and storage of encrypted survey results. Thereby a lot of meta-data is accumulated, such as which IP address participated in which study. Due to easy usability for participants through a browser, we are susceptible to malicious JavaScript files exfiltrating answers to researchers directly or interested third parties.
WAIT~\cite{Meissner2021} showcases an extension for web browsers that enables the verification of provided web applications against a public transparency log, similar to certificate transparency.
Approaches like this, further limit the attack possibilities of the platform provider and may be applied to multiple similar scenarios to alleviate this limitation.
Another possible attack vector is a compromise of the enclave by breaking the sandbox of the JavaScript engine via the statistical analysis script.

%% file: sections/5-relatedwork.tex
%!TeX spellcheck = en-US
\section{Related Work}
\label{sec:relatedwork}

Quite recently, researchers from psychology (\eg Peikert et al.~\cite{Peikert2019}, Arslan et al.~\cite{Arslan2020}) have started to suggest standardized and partially automated workflows to support their research process based on open source technologies. 
They primarily take into account reproducibility, replicability, and more transparency for the research process, but their workflows are also designed to streamline research efforts.
However, they have neither advanced privacy protection mechanisms for collected data, nor do they comprehensively protect the integrity of the study design and the data analysis.

Lapets et al.~\cite{Lapets2016} suggest a system for secure analytics that can be deployed as a web application.
Data contributors add a random mask to their data before uploading it to a service provider, which aggregates all data entries.
The mask is encrypted using the public key of an analyzer and also uploaded.
The analyzer can retrieve the aggregated masked data and the encrypted masks, which can be decrypted, aggregated, and subtracted from the aggregated masked data to retrieve the true aggregate data.

Rmind~\cite{Bogdanov2018} is a system for privacy-preserving statistical analysis based on secure multi-party computation.
Input data is secret shared across three computation servers that are assumed to be non-colluding.
Statistical queries can be evaluated using an R-like statistics language.

STAR~\cite{Schreiber2019} enables statistical tests with auditable results based on secret sharing, secure multiparty computation, and tamper-proof ledgers.
The main focus of STAR is to address the issue of false discoveries in scientific studies, by generating a publicly verifiable certificate for the performed statistical tests.

Lu et al.~\cite{Lu2016} designed a system to enable privacy-preserving statistical analysis with fully homomorphic encryption. While they proposed protocols for multiple statistics, performance is inferior and the privacy policies by the ethics boards cannot be enforced.

The above systems all assume non-colluding computation parties.
However, for quantitative empirical studies this assumption does not hold, because the computing parties have no incentive not to collude, as they are disjunct from the data subjects.
In case the computing parties are research institutions, they may even have a strong incentive to collude to gain the raw data for exploration.
In contrast, \peqes embraces and extends the existing research workflow and integrates ethics committees to review studies for privacy risks.

%% file: sections/6-discussion.tex
%!TeX spellcheck = en-US
\section{Discussion and Future Work}
\label{sec:discussion}

The practicality of our platform depends essentially on the acceptance by the researchers and the integratability of technical steps into existing organizational procedures.

We assume that researchers might perceive the platform rather skeptically at first.
After successful submission of a study to the platform, the subsequent study progress\,---\,\ie data collection and data analysis\,---\,becomes neither observable nor fully controllable by them.
The platform appears more like a black box that ingests participations and eventually yields the output of a pre-defined analysis.
The platform also takes away the possibility of interactive data exploration from the researchers.
Furthermore, it prevents any post-hoc modifications of statistical analyses.
However, we argue that these restrictions are in line with the concept of pre-registration and encourage good scientific practices.
The workflow deliberately decouples hypothesis generation and hypothesis testing, data collection and data analysis, as well as study design and study execution.
Adhering to this approach ensures a rigorous conduct and thus increases the trust in its outcomes.
There are indeed methods to circumvent negative implications of this approach when conducting studies in practice.
Before implementing comprehensive confirmatory studies, researchers typically apply purely explorative analyses to generate hypothesis ideas and rely on pilot studies to test operationalizations, inventories, and surveys.
So in such a two-step approach, \peqes will only be used in the second step for the larger, confirmatory study, but not for the initial exploration.
Regarding the statistical analysis scripts, we argue that such scripts should be tested in advance against synthetically generated survey responses.
This is not only helpful for development and debugging of the analysis code. 
A prior execution with simulated and controllable responses for test purposes also helps to verify the correctness of the intended analysis steps in the first place. 

Note that the analysis scripts can include data wrangling and data cleansing steps as part of the overall analysis.
These steps are therefore also subject to approval by the ethics board.
This also allows researchers to apply tests for certain assumptions to their samples (\eg Kolmogorov-Smirnov test for normality).
To include such pre-tests is particularly relevant for non-interactive data analysis procedures as potential biases in the samples cannot be observed exploratively anymore.  
By performing appropriate tests, biased samples can either be excluded, or the data analysis process would stop its execution as soon as the distribution assumptions for the sample do not hold.

Regarding the integration into existing procedures, we assume that the review step of ethics board represents the main barrier.
Whereas until now ethics boards only review a written description of a study, the new workflow requires them to audit and sign a structured study specification and to assess the privacy impact of a designated data analysis and its estimated output.
Again, we believe that this change eventually strengthens the position of reviewing boards, as it increases compliance and strengthens their influence on the actual study execution.
A possible step towards the simplification of this procedure is the specification and enforcement of data policies.
Here, the system would derive required data usage policies from the study specification.
In turn, the ethics board would review these policies and compare them with general guidelines.
Ultimately, the platform would prevent any usage of participants' data that is not in line with the pre-defined policies or that interferes with specific policies configured by a participant.
An orthogonal approach would be to not only require the review of the ethics board, but additionally necessitate peer reviews of the pre-registered study design.
Peers in the research community can be expected to have more domain expertise for the specific research and study design and can complement the judgment of the ethics board.

\peqes and its workflow is also advantageous for the reproducibility and replicability of studies.
Reproducibility assures that the same original data with the same statistical analysis code reproduce the same findings.
This is inherently provided by \peqes, as the original data cannot be changed, and the execution of the statistical analysis is also taking place within the protected  boundaries of \peqes.
Replicability describes a repetition of a study with the same methods, but with newly collected original data and independent from the original research efforts. 
In this case, the published study design artifacts can be used to exactly replicate an existing study at another institution that runs a \peqes instance. 
This will require an attestation of the ethics board which is responsible for the replicating instance.
However, \peqes guarantees that the exact same methods are used during the study replication.
In this way, \peqes excludes confounding factors during the replication and makes it possible to attribute the replicated results solely to the new samples.

The success of \peqes ultimately depends on the degree to which it can be extended into a comprehensive solution for privacy-enhanced empirical research.
Our next step is the design of a privacy-enhanced participant management component to tackle longitudinal studies with recurrent participations, attribute-based participant recruitment (\eg only targeting left-handed subjects in a certain group), and reward payoffs\,---\,in each case with appropriate privacy protection.
Inclusion of such a system can also further strengthen the authenticity of participants' data by limiting the possibility of sybil attacks by participants and creation of fake participants by malicious researchers.
Data access policies for analysis scripts and data usage policies of study responses would also facilitate secondary data usage.
\Eg researchers could submit supplementary studies that operate on the same data sets, provided that the ethics board has approved the supplementary analysis and provided that the participant has agreed to secondary use of their data in the first place.

For the future, we also expect our platform to handle data collections beyond online surveys.
This includes scenarios that are inherently more sensitive to data privacy issues, such as mobile sensing.
\Eg the use of apps to collect individual behavioral patterns of participants throughout the day makes strong technical data protection imperative.

%% file: sections/7-conclusion.tex
%!TeX spellcheck = en-US
%
\section{Conclusion}
\label{sec:conclusion}

We have suggested a privacy-enhanced workflow for quantitative empirical studies that embraces the idea of pre-registration and adds additional steps to enforce the correct workflow execution.
Furthermore, we have introduced \peqes, a trusted execution platform that implements this workflow based on appropriate protocols and Intel SGX.
Our solution addresses the orderly execution of a pre-defined study design, guarantees the confidentiality of participants' data, ensures the integrity of the study and its analysis procedures, and thus increases trust in corresponding outcomes and results.
The evaluation of the \peqes prototype showed a noticeable performance overhead of our solution, which in our opinion is still negligible in practice.
\peqes represents a core building block towards a more comprehensive technical solution that improves privacy protection for quantitative empirical studies by adding complementary components such as participation management.